\documentclass[11pt]{article}
\usepackage{amsmath,amsbsy,amsfonts,amssymb,graphics,epsfig,float,mathrsfs,amsthm,braket}
\usepackage{ifsym}
\usepackage{psfrag}
\usepackage{color}
\usepackage[normalem]{ulem}

\newcommand \beq{\begin{equation}}
\newcommand \eeq{\end{equation}}

\newcommand{\Sqq}{\bar{\sigma}_{\theta\theta}}
\newcommand{\Srr}{\bar{\sigma}_{rr}}
\newcommand{\Srq}{\bar{\sigma}_{r\theta}}
\newcommand{\sqq}{\sigma_{\theta\theta}}
\newcommand{\srr}{\sigma_{rr}}

\newcommand{\lp}{\ell_{p}}
\newcommand{\upd}{\mathrm{d}}

\newcommand{\Kcurv}{K_{\mathrm{curv}}}
\newcommand{\Ktens}{K_{\mathrm{tens}}}
\newcommand{\medge}{m_{\mathrm{edge}}}
\newcommand{\srq}{\sigma_{r\theta}}
\newcommand{\cF}{{\cal F}}
\newcommand{\tdelta}{\tilde{\delta}}
\newcommand{\sminp}{\sigma_{\mathrm{min}}^{(p)}}

\newcommand{\s}[1]{{\textsf{\textbf{#1}}}}

\begin{document}
\title{\s{Regimes of wrinkling in pressurized elastic shells}}
\author{ \textsf{Matteo Taffetani$^\dagger$ and Dominic Vella$^\dagger$}\\ 
{\it$^\dagger$Mathematical Institute, University of Oxford, UK}}

\date{\today}
\maketitle
\hrule\vskip 6pt
\begin{abstract}
We consider the point-indentation of a pressurized elastic shell. It has previously been shown that such a shell is subject to a wrinkling instability as the indentation depth is quasi-statically increased. Here we present detailed analysis of this wrinkling instability using a  combination of analytical techniques and finite element simulations. In particular, we study how the number of wrinkles observed at the onset of instability grows with increasing pressurization. We also study how, for fixed pressurization, the number of wrinkles changes both spatially and with increasing indentation depth beyond onset. This `Far from threshold' analysis exploits the largeness of the wrinkle wavenumber that is observed at high pressurization and leads to quantitative differences with the standard `Near threshold' stability analysis.
\end{abstract}
\vskip 6pt
\hrule

\maketitle

\section{Introduction}

The buckling instability of elastic objects is a staple of much research in engineering, applied mathematics and physics. Traditionally, the focus has been on determining the properties of these instabilities close to the onset of instability with a view to avoiding them: where does instability occur and how sensitive to imperfections is this threshold \cite{Budiansky1966,Budiansky1979}? However, more recently this `buckliphobia' has, to a certain extent, been replaced by `buckliphilia' \cite{Reis2015}: the patterns that are commonly observed in instability (notably wrinkling, folding and creasing) are  not only common in the natural world (for example in skin \cite{Ciarletta2013_qjmam}, the cortical convolutions in the human brain \cite{Tallinen2016_np} and in growing plants \cite{Li2012}) but are also useful in generating regular patterns in technological applications including photonics \cite{Kim2012} and stretchable electronics \cite{Rogers2010}. In these applications, interest is focussed not so much on the onset of instability, but rather the behaviour far from onset, where novel behaviour,  such as period doubling \cite{Brau2011_np,Cao2012}, may emerge. 

A classic example of elastic instability is the wrinkling of a thin elastic object (a beam) under compression on a liquid bath  or elastic foundation  \cite{Brau2011_np, Pociavavsek2008_s,Fu2015_prsa,Ciarletta2013_prl}. As well as being visually  striking, these wrinkle patterns allow, in principle, for a simple assay of a sheet's mechanical properties \cite{Huang2007}. The essential mechanics of the pattern selection are clear: there is a competition between destabilizing effects (here a compression) and stabilizing effects, such as a bending stiffness (as in Euler buckling). However, the selection of an intermediate length scale, different from the system size, requires a competition between restoring forces. For example, in the compression of an elastic beam (of bending stiffness $B$) on a foundation (of stiffness $K$), the resistance to bending is minimized through the selection of the largest wavelength possible while the foundation prefers the smallest wavelength possible. The wavelength that is observed emerges from a compromise between these two \cite{Cerda2003_prl}; in particular
\beq
\lambda=2\pi \left(\frac{B}{K}\right)^{1/4}.
\label{eqn:simplambda}
\eeq

In uniaxial compression, the result \eqref{eqn:simplambda} applies both at the onset of instability and far beyond it. However, in other scenarios, the behaviour of the system changes qualitatively beyond onset. For example, attempts to determine the location of wrinkles based on determining where compression would occur ordinarily under-predict the wrinkle extent dramatically \cite{Huang2007,Vella2010,Vella2011_prl}. It is now well known that to correctly predict the location of wrinkles it is necessary to account for the fact that wrinkling relaxes the compressive stress that caused instability in the first place \cite{Vella2015_epl,Schroll2013_prl} --- what is usually referred to as tension-field theory \cite{Steigmann1990_prsa}, or the relaxed energy approach \cite{Pipkin1986_jam}. However, an understanding of how this stress relaxation affects the wrinkle wavelength has only come more recently \cite{Cerda2003_prl, Davidovitch2011,Paulsen2016_pnas}. In particular, Cerda \& Mahadevan \cite{Cerda2003_prl} showed that tension along the direction of the wrinkles gives rise to an effective `tensional' stiffness, $\Ktens$, that may be substituted into \eqref{eqn:simplambda} in place of $K$. Generalizing this work, Paulsen \emph{et al.}~\cite{Paulsen2016_pnas} recently showed that curvature along the wrinkle direction can also give rise to a `curvature-induced' stiffness, $\Kcurv$.

In this paper we give a new perspective on wrinkling `Far from threshold' (FT) for the indentation-induced wrinkling of a pressurized elastic shell \cite{Vella2011_prl,Vella2015_epl}. We use both asymptotic analysis of the shallow shell equations, together with ABAQUS simulations. In particular, we perform the FT expansion \cite{Davidovitch2011} by exploiting the large number of wrinkles, $m\gg1$, that are frequently observed in such scenarios, and determine results for the wrinkle number $m$ as  the indentation depth, pressurization and also radial position within the shell vary. This also allows us to generalize the curvature-induced stiffness \cite{Paulsen2016_pnas} to the case in which the object has a natural curvature.

A key theme throughout this work will be comparing and contrasting the results at the onset of instability, or `Near Threshold', with those `Far from Threshold'. Indeed, our study is the first that observes the transition from Near Threshold to Far from Threshold in direct numerical simulations. We therefore begin with a discussion of the governing shallow shell equations in \S\ref{sec:shallow} before then performing the standard `Near Threshold' analysis in \S\ref{sec:NT}. In \S\ref{SEC:FinitElementSimulation} we present our ABAQUS simulations, which motivate the `Far from Threshold' analysis presented in \S\ref{sec:FT}. We compare these results in \S\ref{sec:compare} before summarizing our results in \S\ref{sec:conclude}.

\section{Shallow shell formulation\label{sec:shallow}}

\subsection{Governing equations}

\begin{figure}[!h]
\centering\includegraphics[width=5.0in]{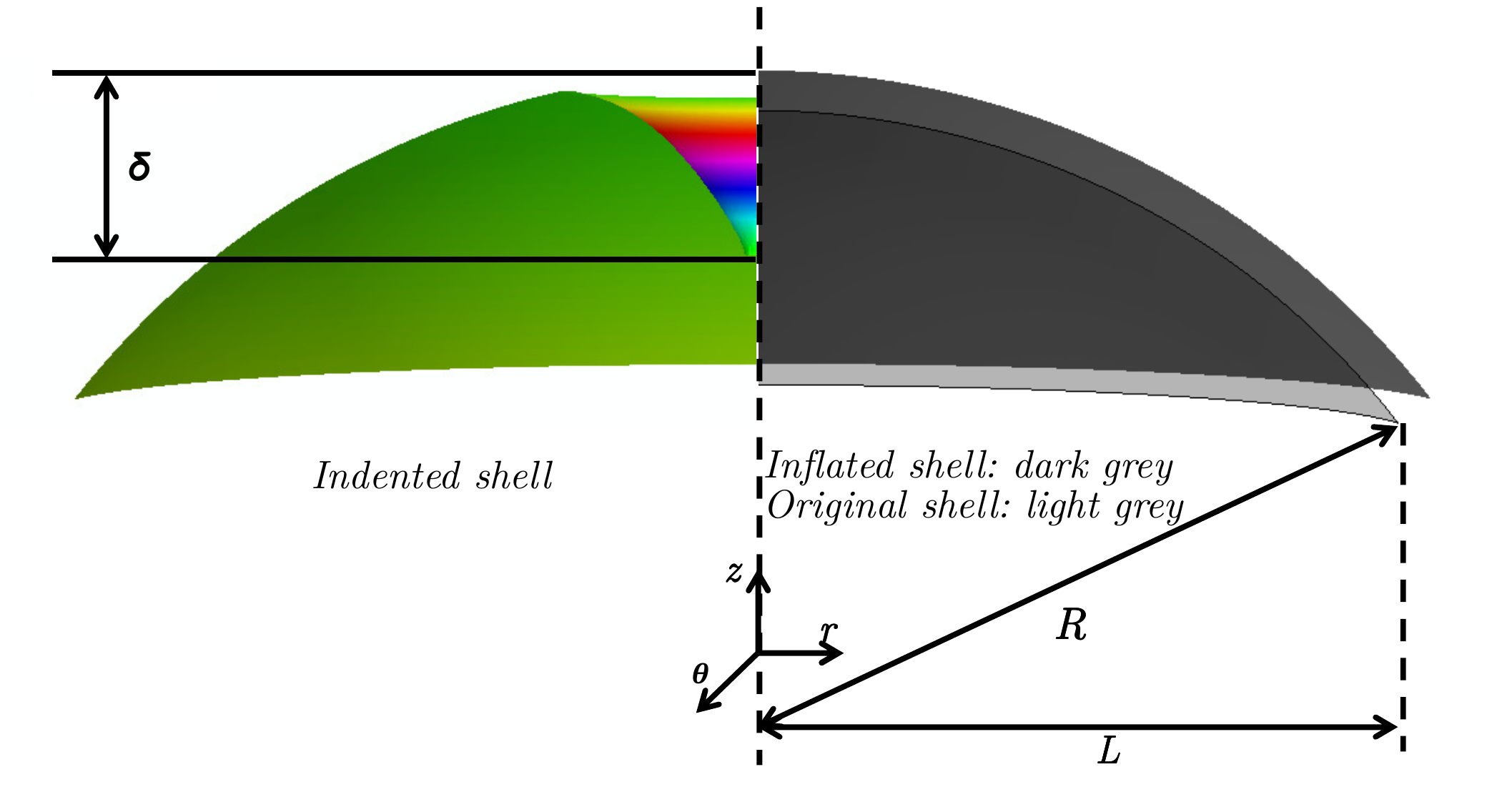}
\caption{ Schematic of a spherical elastic shell subject to an internal pressure $p$ and a point indentation. The original (unpressurized) shell (light grey)  is first inflated to a prescribed internal pressure, $p$, (dark grey). A point indentation, of depth $\delta$, is then applied at the shell's apex.}
\label{FIG:setup}
\end{figure}

Figure \ref{FIG:setup} shows a schematic of a  pressurized, spherical elastic shell, subject to an indentation at its pole. Provided that the horizontal lengthscale over which the shell is deformed by indentation is much smaller than the radius of curvature of the shell, $R$, a convenient framework within which to analyse the deformations of such thin elastic shells is `shallow-shell theory' \cite{VentselKrauthammer,Vella2011_prl}: the intrinsic geometry of the shell is assumed to be identical to the geometry of the plane of its projection. We shall discuss the circumstances under which this assumption is valid later, but for now we introduce a cylindrical polar coordinate system $(r, \theta, z)$ and orient the shell with its planform lying  in the $(r, \theta)$ plane. The key quantities of interest are the deflection of the shell in the normal direction, $w(r,\theta)$, and the stress components within the shell, $\srr,\sqq$ and $\srq$. For the shell to be in equilibrium, the in-plane stress balances, i.e.
\beq\label{EQ:equilibrium}
\frac{1}{r}\frac{\partial }{\partial r}\left(r \sigma_{rr}\right)-\frac{\sigma_{\theta\theta}}{r} + \frac{1}{r}\frac{\partial \sigma_{r\theta}}{\partial \theta}=0,\quad
\frac{1}{r}\frac{\partial }{\partial r}\left(r \sigma_{r\theta}\right)+\frac{\sigma_{r\theta}}{r} + \frac{1}{r}\frac{\partial \sigma_{\theta\theta}}{\partial \theta}=0,
\eeq must be satisfied, together with the normal force balance. For a spherical elastic shell of radius $R$, subject to a constant internal pressure $p$ and a localized indentation force $F$, this normal force balance reads \cite{VentselKrauthammer,Vella2011_prl,Vella2012}
\begin{align} \label{EQ:force_balance_stress}\begin{split}
B\nabla^2 \nabla^2 w +\frac{ \srr+\sqq}{R}&-\srr\frac{\partial^2 w}{\partial r^2} - \sqq\left(\frac{1}{r}\frac{\partial w}{\partial r} + \frac{1}{r^2}\frac{\partial^2 w}{\partial \theta^2}\right) \\
&+ 2\sigma_{r \theta}\left(\frac{1}{r}\frac{\partial^2 w}{\partial r \partial \theta} - \frac{1}{r^2}\frac{\partial w}{\partial \theta}\right) = p-\frac{F}{2\pi}\frac{\delta(r)}{r}.
\end{split}\end{align} Here $B=Et^3/[12(1-\nu^2)]$ is the bending stiffness, with $E$ the Young modulus, $t$ the thickness and $\nu$ the Poisson ratio of the shell. The localized force $F$ is implemented as a Dirac $\delta$-function but may also be understood in the first integral of \eqref{EQ:force_balance_stress} as a shear force $Q_r=-F/(2\pi r)$ present only outside a cylindrical region of radius $r=\epsilon\to0$ \cite{VentselKrauthammer}. Note that in \eqref{EQ:force_balance_stress} the Laplacian operator is that used in plane polar coordinates, i.e.
\begin{align}
\nabla^2 f = \frac{1}{r} \frac{\partial}{\partial r}\left(r\frac{\partial f}{\partial r}\right)+\frac{1}{r^2}\frac{\partial^2 f}{\partial \theta^2}. \notag
\end{align} 

The problem is  closed by combining the linear stress-strain relationships (Hooke's law) with the geometrically non-linear relationships between strain and the in-plane and normal displacements, $\mathbf{u}=(u_r,u_\theta)$ and $w(r,\theta)$. For a spherical shell, these read
\begin{align}\label{EQ:err_dim}
\frac{\partial u_r}{\partial r} + \frac{1}{2} \left(\frac{\partial w}{\partial r}\right)^2 + \frac{w}{R}=\epsilon_{rr} = \frac{1}{Et} \left(\srr - \nu \sqq\right),
\end{align}
\begin{align}\label{EQ:ett_dim}
\frac{u_r}{r} + \frac{1}{r} \frac{\partial u_{\theta}}{\partial \theta} + \frac{1}{2 r^2} \left(\frac{\partial w}{\partial \theta}\right)^2 + \frac{w}{R} =\epsilon_{\theta \theta} = \frac{1}{Et}\left(\sqq - \nu \srr\right),
\end{align} and
\begin{align}\label{EQ:ert_dim}
\frac{1}{2 r} \frac{\partial u_r}{\partial \theta} +\frac{1}{2} \frac{\partial u_{\theta}}{\partial r} - \frac{u_{\theta}}{2 r} + \frac{1}{2 r} \frac{\partial w}{\partial r} \frac{\partial w}{\partial \theta}=\epsilon_{r \theta}= \frac{1+\nu}{Et}\sigma_{r \theta}.
\end{align}
(Here $\epsilon_{ij}$, with $i,j=r,\theta$ denotes the entries of the symmetric strain tensor.)
We note that in many problems, it is often convenient to introduce an Airy stress function $\Phi$ to ensure the in-plane stresses automatically balance; this then requires the solution of an equation for $\Phi$ that ensures the compatibility of strains. This approach is less convenient for the perturbation analysis that we shall undertake here.

The problem is to be solved subject to appropriate boundary conditions. At the point of indentation, we require that the displacement takes an imposed value, $\delta$, that the shell does not have a cusp, and that the radial displacement $u_r$ vanishes, i.e.
\beq
w(0,\theta) = -\delta, \quad\left.\frac{\partial w}{\partial r}\right|_{r=0} =0,\quad0=u_r(0)=\lim_{r\rightarrow 0} \left[r (\sqq - \nu \srr)\right].
\eeq  
Note also that, for convenience, we imagine imposing a given indentation depth, $\delta$; the force $F$ required to impose that displacement can, in principle, be determined as part of the solution, though we do not discuss the behaviour of this force here.

Far from the indenter, we require the shell to return to its undeformed state. We measure normal displacements from this state so that
\beq
w,\frac{\partial w}{\partial r} \to 0 \quad\mathrm{as}\quad r\rightarrow \infty.
\eeq We also note that the pressure jump across the shell requires a uniform, isotropic tension within the shell $\srr=\sqq=pR/2$ prior to indentation (by analogy with Laplace's law). We therefore also require that
\beq
 \srr,\sqq \sim \frac{pR}{2}\quad\mathrm{as}\quad r\rightarrow \infty.
\eeq

\subsection{Non-dimensionalization}

In previous studies \cite{Vella2011_prl,Vella2015_epl} it has proved useful to introduce the shell `capillary length'
\beq
\lp = \left(\frac{p R^3}{E t}\right)^{1/2}
\eeq as a natural horizontal length scale. (The shell `capillary length' is the horizontal distance over which the pressure-induced tension, $\sim pR$, together with a restoring force from the curvature of the shell, $Et/R^2$, return the shell to flat following a small vertical deformation; it is somewhat analogous to the capillary length of a liquid--vapour interface.) By then introducing dimensionless variables
\begin{align}\label{EQ:nondimensionalization}
\rho=\frac{r}{\lp},\quad W(\rho,\theta)=\frac{R}{\lp^2}w(r,\theta),\quad\bar{\sigma}_{ij} = \frac{\sigma_{ij}}{p R},\quad \bar{\epsilon}_{ij} = \frac{R^2}{\lp^2}\epsilon_{ij},\quad U_r=\frac{R^2}{l_p^3} u_r,\quad U_{\theta}=\frac{R^2}{l_p^3} u_{\theta}.
\end{align} 
With this non-dimensionalization, the normal force balance equation \eqref{EQ:force_balance_stress} becomes
\begin{align}
\tau^{-2}\nabla^2 \nabla^2 W + \Srr+\Sqq &-\Srr \frac{\partial^2 W}{\partial \rho^2} -\Sqq\left(\frac{1}{\rho}\frac{\partial W}{\partial \rho} + \frac{1}{\rho^2}\frac{\partial^2 W}{\partial \theta^2}\right)\nonumber \\
&+ 2\Srq\left(\frac{1}{\rho}\frac{\partial^2 W}{\partial \rho \partial \theta} - \frac{1}{\rho^2}\frac{\partial W}{\partial \theta}\right) = 1-\frac{\cF}{2\pi}\frac{\delta(\rho)}{\rho}\label{EQ:Vertical_nondimensionalized}
\end{align}
where $\cF=F/(p \lp^2)$ is the dimensionless indentation force and
\beq
\tau = \frac{p R^2}{(E t B)^{1/2}},
\eeq  is the dimensionless pressure. Alternatively, $\tau^{-2}$ is the dimensionless bending stiffness, so that  $\tau^2$ may be thought of as the `bendability' of the shell \cite{Davidovitch2011,Hohlfeld2015_pre}. (Note now that the operator $\nabla$ is written in terms of the dimensionless radius $\rho$ in the obvious way.)
The dimensionless in-plane stress balances become
 \begin{align}\label{EQ:Radial_nondimensionalized}
\frac{1}{\rho}\frac{\partial }{\partial \rho}\left(\rho \Srr\right)-\frac{\Sqq}{\rho} + \frac{1}{\rho}\frac{\partial \Srq}{\partial \theta}=0
\end{align} 
\begin{align}\label{EQ:Azimuthal_nondimensionalized}
\frac{1}{\rho}\frac{\partial }{\partial \rho}\left(\rho \Srq\right)+\frac{\Srq}{\rho} + \frac{1}{\rho}\frac{\partial \Sqq}{\partial \theta}=0,
\end{align} 
while the stress-strain-displacement relations become
\begin{align}\label{EQ:err}
\frac{\partial U_r}{\partial \rho} + \frac{1}{2} \left(\frac{\partial W}{\partial \rho}\right)^2 + W=\bar{\epsilon}_{rr} = \Srr - \nu \Sqq
\end{align}
\begin{align}\label{EQ:ett}
\frac{U_r}{\rho} + \frac{1}{\rho} \frac{\partial U_{\theta}}{\partial \theta} + \frac{1}{2 \rho^2} \left(\frac{\partial W}{\partial \theta}\right)^2 + W =\bar{\epsilon}_{\theta \theta} = \Sqq - \nu \Srr
\end{align}
\begin{align}\label{EQ:ert}
\frac{1}{2 \rho} \frac{\partial U_r}{\partial \theta} +\frac{1}{2} \frac{\partial U_{\theta}}{\partial \rho} - \frac{U_{\theta}}{2 \rho} + \frac{1}{2 \rho} \frac{\partial W}{\partial \rho} \frac{\partial W}{\partial \theta}=\bar{\epsilon}_{r \theta}= (1+\nu)\bar{\sigma}_{r \theta}.
\end{align}

The non-dimensionalization also introduces a second dimensionless parameter,
\beq
\tdelta=\delta R/\lp^2,
\eeq which is the dimensionless indentation depth. 
Generally, we shall be interested in the limit of highly pressurized (and hence highly bendable), and highly indented shells, i.e.~$\tau,\tdelta\gg1$. These limits may seem to be in contradiction with the additional requirements that the typical strains be small, $pR/Et\ll1$, (for Hooke's law to remain valid) and that the lateral scale of deformation $\lp\ll R$ (for the shallow shell theory to be valid). For a given indentation depth, shell radius and elastic modulus, these restrictions can in fact be satisfied in the limit of vanishing pressure and thickness, $p,t\to0$, provided that $p/t\to0$ but $p/t^2\to\infty$, as discussed elsewhere \cite{Vella2015_epl}.

\subsection{Axisymmetric behaviour}

\begin{figure}
\centering
\includegraphics[width=12cm]{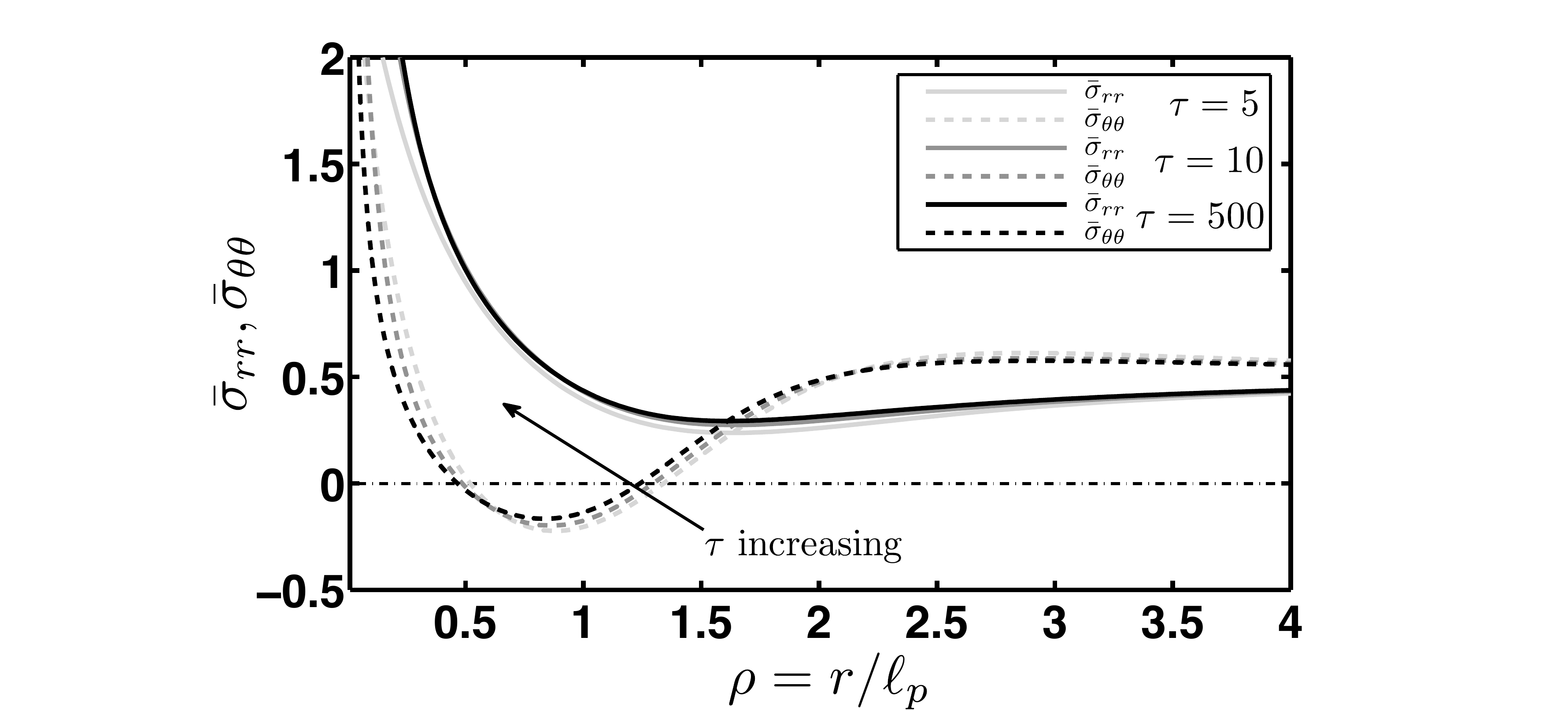}
\caption{Dimensionless stress profiles at a fixed dimensionless indentation depth $\tdelta=4$, with the value of the dimensionless pressure $\tau$ increasing. Note that as $\tau$ increases, the stresses approach a well-defined limit (given by membrane theory) and that, for this value of $\tdelta$, the hoop stress is compressive in some annular region, $\Sqq<0$ for $0.5\lesssim\rho\lesssim1.3 $. (The line $\Sqq=0$ is shown for clarity by a horizontal dash-dotted line.) In highly bendable shells, $\tau\gg1$, this compression leads to wrinkling, which is the focus of this paper. }
\label{fig:stressprofs}
\end{figure}

Axisymmetric solutions of \eqref{EQ:Vertical_nondimensionalized}--\eqref{EQ:ert} have been studied in some detail \cite{Vella2011_prl,Vella2012}. The key observation is that as $\tdelta$ increases,  the hoop stress $\Sqq$ is non-monotonic: close to the indenter the stress becomes large, while far away it takes its undisturbed far-field value, $\Sqq\sim 1/2$. Between these two tensile regions, however, there is an annulus in which the hoop stress becomes relatively compressive; this behaviour can be understood as follows: indentation acts to pull material circles in to a position where their circumference is too long for their final position and so they become relatively compressed. As $\tdelta$ grows, this hoop stress becomes more and more compressive (relative to the far-field value, $\sqq\to1/2$) until eventually it becomes absolutely compressive, $\Sqq<0$. In figure~\ref{fig:stressprofs} we see that this transition has already occurred by $\tdelta=4$: as $\tau$ increases, the hoop stress $\Sqq<0$ in some region, even as $\tau\to\infty$. A more detailed calculation shows that with $\tau=\infty$ this transition occurs at $\tdelta=\tdelta_c^{\infty}\approx2.5199$ \cite{Vella2011_prl}.

In the limit of large $\tau$ the shell becomes a membrane-shell, i.e.~it has zero resistance to compression and we expect that it should buckle for any $\tdelta>\tdelta_c^{\infty}$. At finite values of $\tau$, the classic problem is to try and understand \emph{when} this buckling first occurs: what is the smallest indentation depth $\tdelta_c(\tau)$ for which a buckled solution exists? What buckling mode is adopted at this buckling point? These questions may be addressed by a standard linear stability analysis, which we refer to as the `Near Threshold' (NT) analysis (since it is only valid very close to the threshold of instability).

\section{`Near threshold' analysis\label{sec:NT}}

To understand the onset of buckling in this problem, we begin by making a small perturbation, with wavenumber $m$, of the axisymmetric state. We therefore take the ansatz
\begin{align}\begin{split}\label{EQ:linear perturbation_dimension}
W\left(\rho,\theta\right)= W^{(0)}\left(\rho\right) + W^{(1)}\left(\rho\right)\cos m \theta\\
\bar{\sigma}_{ij}\left(\rho, \theta\right)= \bar{\sigma}_{ij}^{(0)}\left(\rho\right) + \bar{\sigma}_{ij}^{(1)}\left(\rho\right)\cos m \theta
\end{split}
\end{align}
where superscript $(0)$ indicates an axisymmetric quantity and a $(1)$ indicates a small perturbation.

We saw above that in the limit $\tau\to\infty$ there is a critical indentation depth, $\tdelta_c^{\infty}\approx2.5199$, at which $\Sqq$ first becomes compressive.  We expect that for $\tau<\infty$ this will be a lower bound on the indentation $\tdelta_c(\tau)$ required to cause buckling. We determine $\tdelta_c(\tau)-\tdelta_c^\infty$ by linearizing the governing equations for the perturbation quantities, i.e.~those with a $(1)$ superscript. The result is a quadratic eigenvalue problem \cite{Tisseur2001} for $m^2$ with a given indentation depth $\tdelta$; this eigenvalue problem does not have a solution for $\tdelta<\tdelta_c(\tau)$, allowing the critical indentation depth, together with the associated mode number, to be determined numerically.

\begin{figure}[h]
\centering\includegraphics[width=13.7cm]{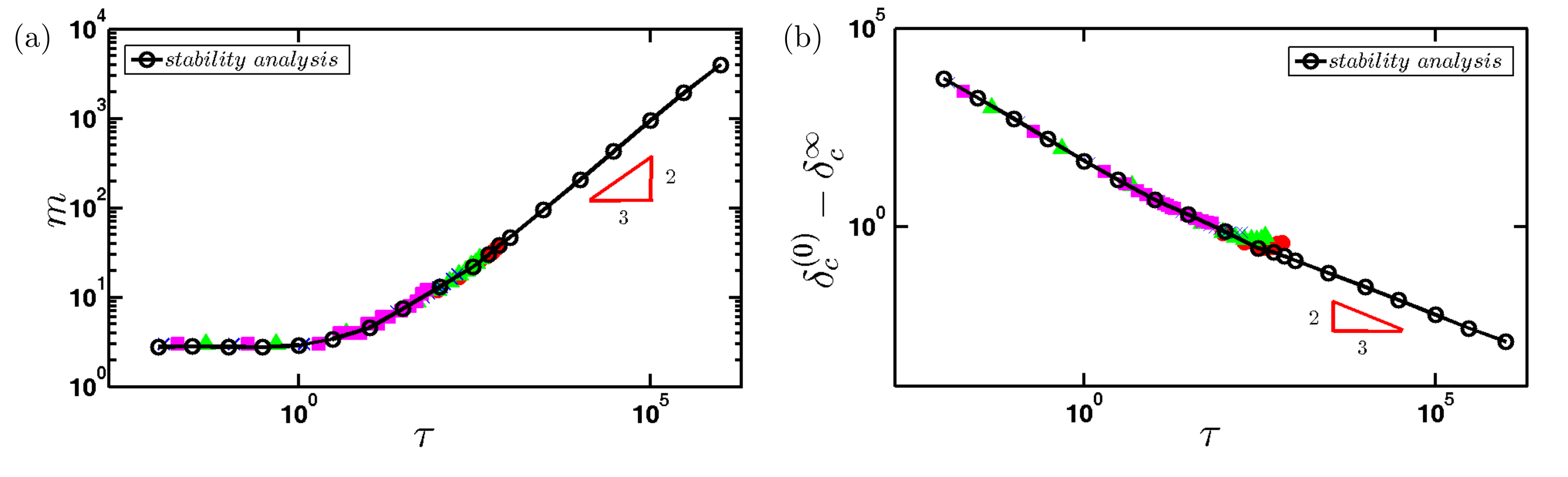}
\caption{Results from the numerical linear stability analysis discussed here (solid curve), together with the results of previous ABAQUS simulations \cite{Vella2011_prl} (coloured markers represent different simulation parameter values, as in \cite{Vella2011_prl}; in  particular, simulations were performed for shells with Young modulus $E=70\mathrm{~GPa}$, radius $R=1\mathrm{~m}$ and thicknesses $t =0.5\mathrm{~mm}$ (circles), $1\mathrm{~mm}$ (triangles), $2\mathrm{~mm}$ (crosses) and $5\mathrm{~mm}$ (squares) with a range of internal pressures \cite{Vella2011_prl}). (a) The wavenumber $m$ and (b) the critical indentation depth $\tdelta_c(\tau) - \delta_c^{\infty}$ at the onset of wrinkling, both presented as functions of the dimensionless pressure $\tau$.  }\label{FIG:ScalingExperimentsNT}
\end{figure}

Our numerical results  are shown in figure~\ref{FIG:ScalingExperimentsNT} and suggest that in the highly pressurized limit, $\tau\gg1$,
\begin{align}\begin{split}
m \sim \tau^{2/3},\quad \tdelta_c(\tau)-\tdelta_c^{\infty} \sim \tau^{-2/3}.
\label{eqn:NTscalings}
\end{split}\end{align} These results can be rationalized by a scaling analysis of the governing equations, along the lines of the analysis presented in \cite{Davidovitch2011}; the details of this scaling calculation are  to be presented elsewhere for a thin sheet floating at a liquid interface \cite{Vella2017}. Here, however, we focus instead on first comparing these results with the results of more detailed numerical simulations, before probing what happens beyond the onset of instability.

\section{Finite Element Analysis}\label{SEC:FinitElementSimulation}

To provide an alternative perspective on the wrinkling of indented, pressurized elastic shells, we used Finite Element simulations, following previous work \cite{Vaziri2008,Vaziri2009,Vella2015_epl}. These simulations were performed using the commercial finite element code ABAQUS 6.14 (Dassault Syst{\'e}mes Simulia Corp., Providence, RI, USA). A linearly elastic (Young's modulus $E= 7 \mathrm{~GPa}$  and Poisson ratio $\nu=0.3$) spherical cap of radius $R=1\mathrm{~m}$, base chord diameter  $2L=1.6\mathrm{~m}$ and thickness $t=0.05\mathrm{~mm}$ is discretized using quadrilateral linear shell elements with reduced integration and finite membrane strain (S4R). The number of elements is chosen to ensure that the resulting mesh is able to properly resolve azimuthal variations with a large wavenumber. Here, we investigate pressurizations up to $\tau=1888$ (for which the number of wrinkles at onset is $m_{\mathrm{NT}}(1888)=70$) in ABAQUS. We therefore ensure that there are enough elements to resolve azimuthal variations with a substantially larger wavenumber; specifically we ensure that our simulations are able to resolve wrinkles with wavenumber $3m_{\mathrm{NT}}(1888)=210$, and find that  94891 elements is sufficient for this purpose. The same number of mesh elements is used in all simulations reported here, and no initial imperfection is imposed. 

Our simulations are conducted in two steps: first, the shell  is inflated to the desired pressure ($p=1 \mathrm{~kPa}$, $p=5 \mathrm{~kPa}$ and $p=10 \mathrm{~kPa}$ are investigated here, corresponding to  $\tau=188$, $\tau=944$ and $\tau=1888$, respectively). This inflation is performed using the built-in procedure *Static, General. The shell is then indented by imposing the desired displacement within a circular area of radius $L_A=5\mathrm{~mm}$ at the apex of the shell; this indentation mimics that caused by a flat punch  whose size still remains much smaller than the shell width $L$ and the typical horizontal length scale $\lp\gtrsim 5\mathrm{~cm}$ though we note that the limit of a point indenter is well-defined and regular \cite{Vella2012}. The indentation step is of the type *Dynamic, Implicit, and stabilizes the solution by introducing a Rayleigh damping $\alpha$ proportional to the mass.  Here the damping coefficient $\alpha$ and the density of the material of the shell are chosen arbitrarily; however, it is verified {\it a posteriori} that the amount of kinetic energy in the simulation remains less than $10\%$ of the strain energy. The simulations may therefore still be considered to be in the {\it quasi-static} regime. Our numerical approach falls within the class of {\it dynamic relaxation techniques} in the classification proposed by Taylor \emph{et al.} \cite{Taylor2015_jmps}: both a mass and a damping (proportional to the mass) are introduced and a dynamic problem is solved. In contrast with the {\it kinetic damping approach} used by Taylor \emph{et al.} \cite{Taylor2014_jmps}, however, we did not optimize the problem (reducing the dynamics to be dependent only on the density parameter).

\begin{figure}[!h]
\centering\includegraphics[width=13cm]{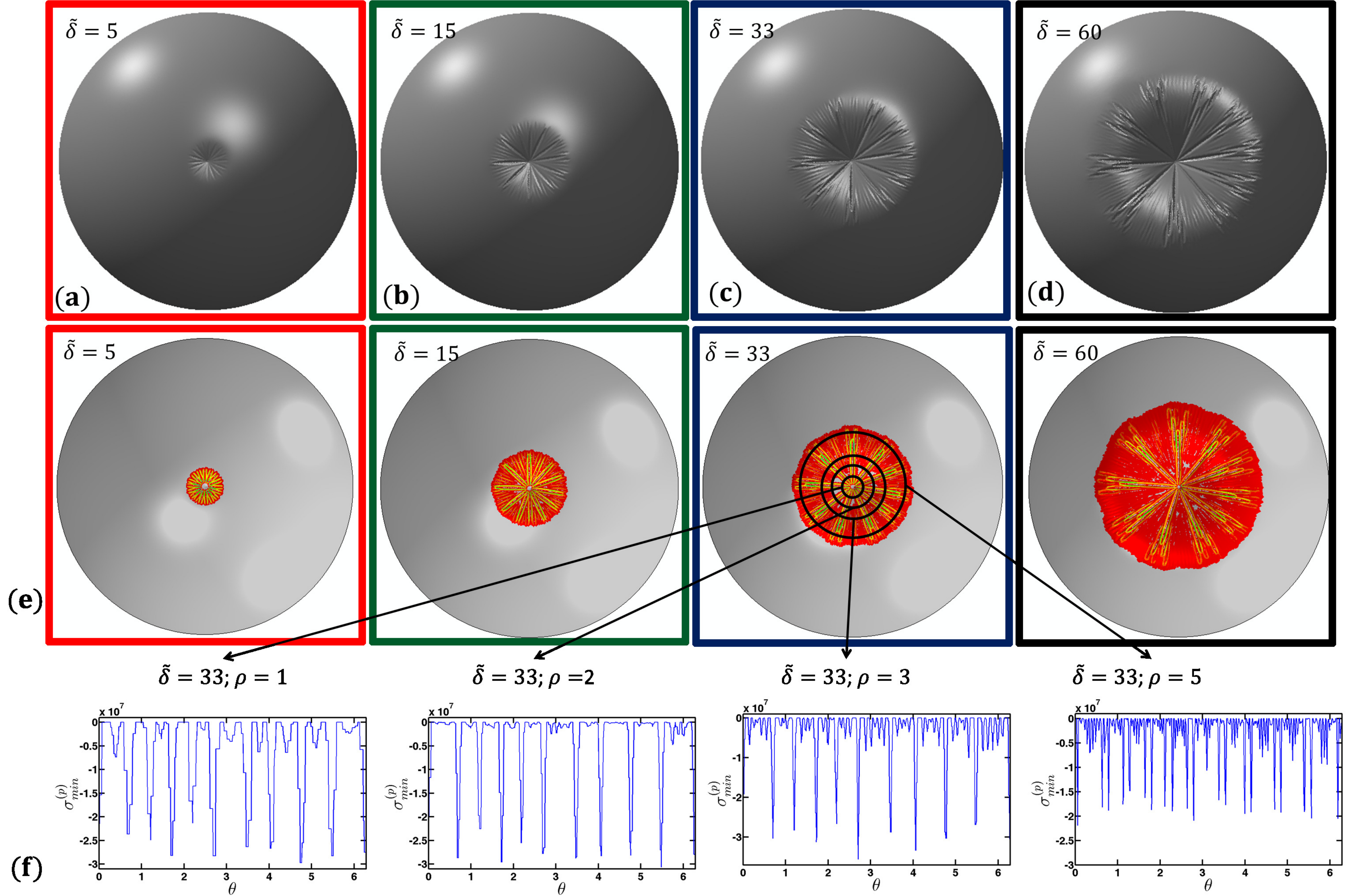}
\caption{Onset and development of wrinkles as determined in ABAQUS simulations with $\tau=188$. (a)-(d) Shell profiles at different indentation depths show the evolution of the wrinkle pattern with indentation depth: (a) $\tdelta=5$, (b) $\tdelta=15$, (c) $\tdelta=33$, and (d) $\tdelta=60$. These shell profiles also show that the wrinkle number varies spatially. (e) The Minimum Principal Stress, $\sminp$, shown as a colour map for increasing indentation depths with $\tau=188$ fixed. Here, colours give a sense of the relative size of $\sminp<0$ with grey denoting $\sminp=0$. (f) The angular dependence of $\sminp$ (in units of $Pa$) at four radial positions when $\tdelta=33$.}
\label{FIG:Deformeds}
\end{figure}

The shapes of a deformed shell with $\tau=188$ held fixed and $\tdelta$ increased, as found in our ABAQUS simulations, are shown in figure~\ref{FIG:Deformeds}(a)-(d).  These images show clearly that the simulations predict a wrinkling instability, as expected. However, two features of these images are immediately inconsistent with the Near Threshold account of wrinkling. Firstly, the pattern evolves with the indentation depth, $\tdelta$. Secondly, the wrinkle number is not spatially uniform: wrinkles appear at different angular positions, depending on the indentation depth (see particularly figure~\ref{FIG:Deformeds}(c), (d)).

Understanding the two features of the wrinkling pattern shown in figure~\ref{FIG:Deformeds}, namely the variation of the wrinkle number in space and with indentation depth, will be our focus for much of the remainder of this paper. However, such a study requires a reliable way of quantifying the number of wrinkles at a given radial position, $\rho$. While the presence of wrinkles in a shell is visually evident in simulations, manually counting wrinkles is neither practical nor reliable. Other methods, including the use of Fourier Transforms give broad ranges of the wrinkle number obscuring the trends evident in images of the shells, see figure~\ref{FIG:Deformeds}(a)-(d), for example. Here, we examine the behaviour of the Minimum Principal Stress (MPS): since wrinkling occurs when the hoop stress becomes sufficiently compressive ($\sqq<0$) and the radial stress remains tensile ($\srr>0$) the MPS should highlight both the direction of the compression and its frequency. Although the normal displacement might be a more natural candidate through which to detect wrinkles, the MPS is expected to vary sinusoidally in the same way. Empirically, we find that the MPS calculated by ABAQUS gives a less noisy signal than the normal displacement, making the identification of the wrinkle wavenumber this way more robust.

Colour maps of the MPS, $\sigma_{min}^{(p)}$ with increasing $\tdelta$ and $\tau = 188$ fixed are shown in figure~\ref{FIG:Deformeds}(e). These plots show that the MPS highlights the wrinkled region: outside the wrinkled region $\sigma_{min}^{(p)}=0$, which is shown in grey. For a given dimensionless indentation depth $\tdelta$, the MPS field at each node of the original mesh is mapped onto a regular polar grid so that the shell may then be cut into slices at increasing radial distance from the indentation point. (This interpolation is performed using the MATLAB function {\it griddata}, equipped with the {\it nearest} method.) Once this has been done, the wavenumber at a given radial distance from the indentation point may be extracted by counting the troughs in the azimuthal variation of the MPS (see figure~\ref{FIG:Deformeds}(f)).

\begin{figure}[!h]
\centering\includegraphics[width=0.8\textwidth]{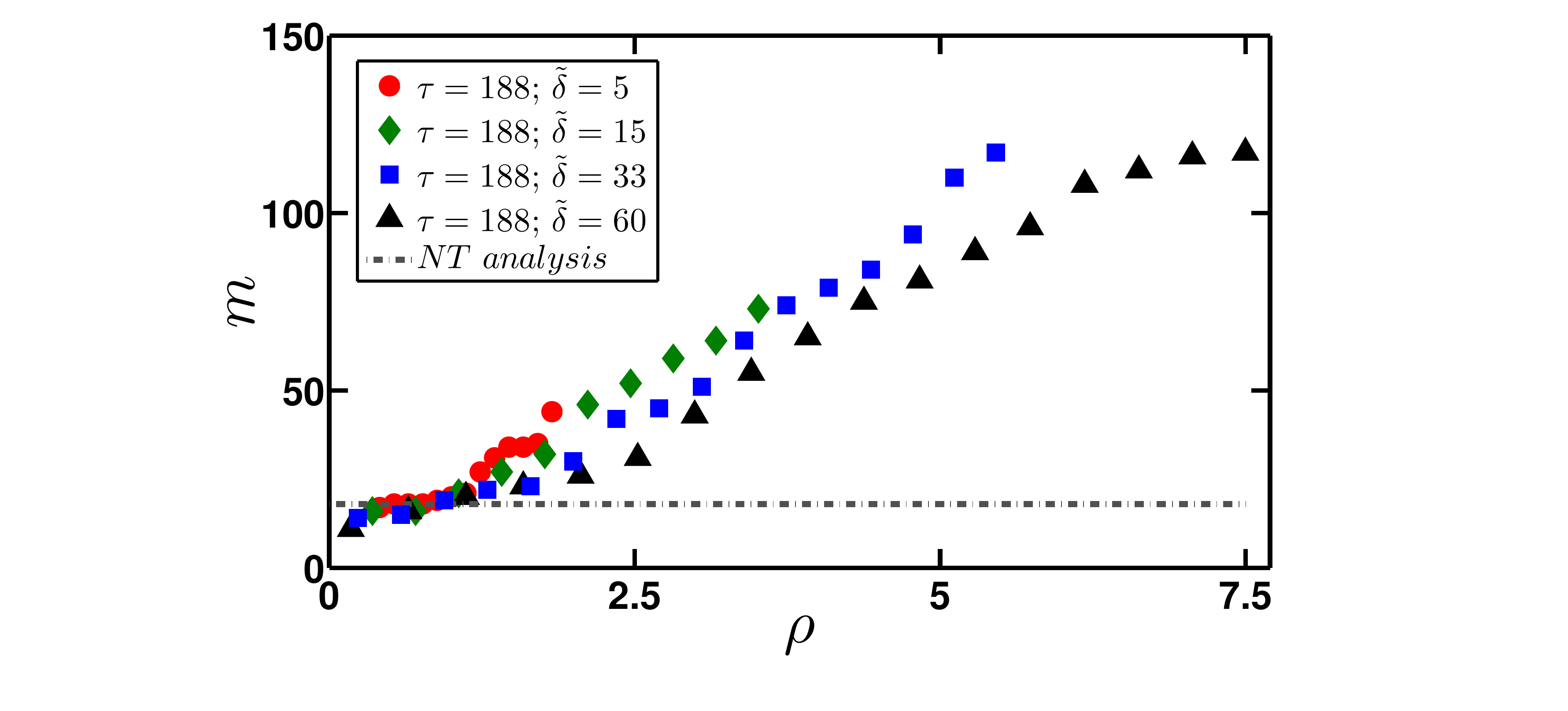}
\caption{The numerically-determined dependence of wrinkle number $m$ on radial distance from the indenter, $\rho$, for ABAQUS simulations with $\tau=188$. The four sets of data correspond to the indentation depths shown in figure~\ref{FIG:Deformeds}(a)-(d): $\tdelta=5$ (circles), $\tdelta=15$ (diamonds), $\tdelta=33$ (squares) and $\tdelta=60$ (triangles). The prediction of the NT analysis ($m_{NT}=18$) is shown as the horizontal dash--dotted line. The means of determining the wrinkle number is discussed in detail in the text.}
\label{FIG:MinPrincipal}
\end{figure}

The raw results of this analysis for the simulations reported in figure~\ref{FIG:Deformeds} are shown in figure~\ref{FIG:MinPrincipal}. As expected based on the images from the simulation, we see  that the wrinkle number varies both with radial position and with indentation depth, $\tdelta$. This is different from the result of the Near Threshold (or linear stability) analysis, which instead predicts a particular wavenumber at the onset of instability. We would like to be able to explain these two features of the patterns quantitatively and note that the wrinkle number $m$, although varying, is typically large, $m\gg1$. We therefore seek a theoretical approach that exploits this largeness to describe the behaviour of the system Far from Threshold (FT).

\section{`Far from Threshold' Analysis \label{sec:FT}}

To understand the evolution of the wrinkle pattern as the indentation depth increases beyond the threshold value $\tdelta_c(\tau)$, we must go beyond the linear stability analysis, and account for the relevant non-linearities. However, to make analytical progress we adapt the approach used by Davidovitch {\it et al.} \cite{Davidovitch2012_pre} and Hohlfeld and Davidovitch \cite{Hohlfeld2015_pre}, to expand the governing equations in terms of a suitable small parameter. In the case of the NT analysis, this small parameter is the wrinkle amplitude and hence the distance beyond the onset of the wrinkling threshold. To understand the behaviour Far from Threshold (an FT analysis), we exploit the observation from the NT analysis that the wrinkling wavenumber $m$ diverges as $\tau \rightarrow \infty$: the parameter $1/m$ seems to be a candidate small parameter.

To make progress, we neglect, for the moment, the observation that the wrinkle number $m$ varies in space and consider a single-mode ansatz for the normal displacement $W(\rho,\theta)$ within the wrinkled zone
\begin{align}\label{EQ:w_expansion}
W(\rho,\theta) = W^{(0)}(\rho) +\frac{1}{m} W^{(1)}(\rho) \cos m \theta.
\end{align} It is then entirely natural to seek expansions for the displacements and components of the stress tensor as 
\begin{align}\label{EQ:field_expansion}\begin{split}
U_{i}(\rho,\theta) = U_{i}^{(0)}(\rho, \theta) + \frac{1}{m}U_{i}^{(1)}(\rho,\theta) + \frac{1}{m^2}U_{i}^{(2)}(\rho,\theta) + O\left(\frac{1}{m^3}\right) \\
\bar{\sigma}_{ij}(\rho,\theta) = \bar{\sigma}_{ij}^{(0)}(\rho, \theta) + \frac{1}{m}\bar{\sigma}_{ij}^{(1)}(\rho,\theta) + \frac{1}{m^2}\bar{\sigma}_{ij}^{(2)}(\rho,\theta) + O\left(\frac{1}{m^3}\right)
\end{split}\end{align}
where superscript $(0)$ indicates the base solution and the superscripts $(1)$ and $(2)$ the higher order corrections. However, we emphasize that the base solutions will not, in general, be solutions of the fully axisymmetric problem \cite{Vella2015_epl}.

The general form of the energy functional in the problem under analysis and in dimensionless variables reads
\beq
{\cal E}_{\mathrm{tot}}= \int\limits_{0}^{2 \pi}\int\limits_{0}^{\infty}  \left\{{\cal E}_{B}  + {\cal E}_{K} + {\cal E}_{H} + {\cal E}_{P} \right\} \rho ~\upd\rho ~\upd\theta
\eeq 
where the terms
\beq\label{EQ:energy_densities}\begin{split}
&{\cal E}_{B} = \tau^{-2}\left\{\left(\nabla^2 W\right)^2 - 2\left(1-\nu\right)\left[\frac{\partial^2 W}{\partial \rho^2} \left(\frac{1}{\rho}\frac{\partial W}{\partial \rho}+\frac{1}{\rho^2}\frac{\partial^2 W}{\partial \theta^2}\right)-\left(\frac{\partial}{\partial \rho}\left(\frac{1}{\rho}\frac{\partial W}{\partial \theta}\right)\right)^2\right]\right\}; \\
&{\cal E}_{K} = \Srr\bar{\epsilon}_{rr} + 2\Srq\bar{\epsilon}_{r\theta}; \quad {\cal E}_{H} = \Sqq\bar{\epsilon}_{\theta \theta}; \quad {\cal E}_{P} = W
\end{split}\eeq
are the bending energy density, the stretching energy density due  to the radial and shear stresses, the stretching energy due to the hoop stress contribution and the work per unit area in compressing the gas, respectively.
The FT approach divides this energy additively as ${\cal E}_{\mathrm{tot}}={\cal E}_{\mathrm{dom}}+{\cal E}_{\mathrm{w}}$ where ${\cal E}_{\mathrm{dom}}$ is the portion of the  system's energy that does not depend on the (small) wrinkle amplitude, and hence dominates the portion that does depend on the wrinkle amplitude, ${\cal E}_{\mathrm{w}}$, or the sub-dominant energy \cite{Davidovitch2011}. In this approach, the macroscopic properties of the wrinkle pattern (such as \emph{where} the shell is wrinkled) is determined by minimizing ${\cal E}_{\mathrm{dom}}$, while the wrinkle number is determined by minimizing ${\cal E}_{\mathrm{w}}$. The position of the wrinkles has been considered in detail before \cite{Vella2011_prl,Vella2015_epl}. Here, we shall instead focus on the number of wrinkles $m$ and so it is interesting to understand the general structure of ${\cal E}_{\mathrm{w}}$, before formally deriving the order by order balances.
Let us insert the expansion given in \eqref{EQ:w_expansion} and \eqref{EQ:field_expansion} and the relations in \eqref{EQ:err}, \eqref{EQ:ett} and \eqref{EQ:ert} into the various energy terms presented in equation \eqref{EQ:energy_densities}. Moreover, let us also anticipate that radial derivatives are small in comparison to azimuthal derivatives, and that, in particular, $\partial/\partial \rho \sim 1$ while $\partial/\partial \theta \sim m$. Since ${\cal E}_{\mathrm{w}}$ is, by definition, the contribution depending on the wrinkle amplitude and considering that the perturbation is sinusoidal, i.e. only the even powers remain after integration in the azimuthal direction, we find that the leading order contributions are
\beq\begin{split}\label{EQ:energy_densities_leading}
& {\cal E}^{LO}_{B}\left(\rho, m\right) = \tau^{-2}\left[\frac{1}{\rho^2} \frac{1}{m}\frac{\partial^2 }{\partial \theta^2}\left(W^{(1)}(\rho)\cos m\theta\right)\right] ^2 = \tau^{-2}m^2\left[W^{(1)}\right]^{2}f_B(\rho)\cos^2m\theta\\
& {\cal E}^{LO}_{K}\left(\rho, m\right) = \frac{1}{m^2}\frac{1}{2}\Srr^{(0)}\left[\frac{\partial }{\partial \rho}\left(W^{(1)}(\rho)\cos m\theta\right)\right]^2 = m^{-2}\left[W^{(1)}\right]^2f_K(\rho)\cos^2m\theta\\
& {\cal E}^{LO}_{H}\left(\rho, m\right) = \frac{1}{m^2}\frac{1}{2 \rho^2} \Sqq^{(0)} \left[\frac{\partial}{\partial \theta}\left(W^{(1)}(\rho)\cos m\theta\right)\right]^2  = \left[W^{(1)}\right]^2f_H(\rho)\sin^2m\theta .
\end{split}\eeq
The most general form of the energy depending on the wrinkle amplitude is therefore
\beq \label{EQ:energy_functional_general}
{\cal E}_{\mathrm{w}}=\pi\int\limits_{0}^{\infty}\left[\tau^{-2}m^2f_B(\rho)+m^{-2}f_K(\rho) +\Sqq^{(0)}f_H(\rho)\right]\left[W^{(1)}\right]^2\rho ~\upd \rho.
\eeq 
Although  how $m$ scales with $\tau$ is not known in advance,  we see from \eqref{EQ:energy_functional_general} that the quantity in the first set of square brackets in the integral is minimized when the terms $\tau^{-2}m^2$ and $m^{-2}$ are of the same order, so that $\tau=O(m^2)$. This is consistent with other Far from Threshold problems in which $m\sim\epsilon^{-1/4}$ with $\epsilon^{-1}$ the bendability, since here $\epsilon^{-1}=\tau^2$ \cite{Cerda2003_prl,Davidovitch2011}. We also note that we expect $\Sqq^{(0)}<0$ (since compression is required for wrinkling); the third term in the integrand of \eqref{EQ:energy_functional_general} therefore decreases the energy of the system,  and hence is the destabilizing term.

\subsection{Expansion of the governing equations}

The expression for the sub-dominant energy leads us to expect that $\tau=O(m^2)$. We therefore now proceed to expand the stress equilibrium equations in \eqref{EQ:Vertical_nondimensionalized}, \eqref{EQ:Radial_nondimensionalized} and \eqref{EQ:Azimuthal_nondimensionalized} together with the stress-strain relations in \eqref{EQ:err}, \eqref{EQ:ett} and \eqref{EQ:ert} order by order in $m^{-1}$, using the expansions given in Equation \eqref{EQ:w_expansion} and \eqref{EQ:field_expansion}. At this stage, we do not know precisely how the wrinkle number $m$ is related to the pressurization $\tau$;
indeed, a key aim of our analysis is to determine more precisely how $m$ and $\tau$ are related. We shall see that the bending stiffness does not enter the first two orders in this Far from Threshold expansion. 

\subsubsection{Order $m$}
The highest order involves only the stresses in the axisymmetric base state (superscript $(0)$), though we shall see that this base state is \emph{not} the same as that determined from the axisymmetric membrane theory. Vertical force balance reveals that
\beq
\frac{1}{\rho^2}\Sqq^{(0)} W^{(1)}\cos m\theta = 0,
\label{EQ:Balance_Om1}
\eeq
while in-plane stress equilibrium gives
\beq
\frac{1}{\rho}\frac{\partial \Srq^{(0)}}{\partial \theta} = 0,\quad\frac{1}{\rho}\frac{\partial \Sqq^{(0)}}{\partial \theta} = 0.
\label{EQ:Balance_Om2}
\eeq
In principle, one could imagine that a term due to bending, $\propto m^2\tau^{-2}$, could enter \eqref{EQ:Balance_Om1}. However, as we have concluded that $\tau=O(m^2)$, we neglect this term at $O(m)$ and see immediately that $\Sqq^{(0)}(\rho,\theta)=0$ (noting that $W^{(1)}(\rho)\neq0$ by the assumption of wrinkling): at leading order the hoop stress is relaxed by wrinkling. This  result is variously known as the tension-field \cite{Steigmann1990_prsa}, or relaxed energy \cite{Pipkin1986_jam} limit, and is well-established as the appropriate limit for wrinkled membranes. We also see from this order that $\Srq^{(0)}(\rho,\theta)=\Srq^{(0)}(\rho)$.

\subsubsection{Order $1$}
Assuming $\Sqq^{(0)}=0$ from the results of the previous order, the normal force balance at $O(1)$ reads
\beq
\label{EQ:Balance_O1A}
\Srr^{(0)}\left(1-\frac{\upd^2 W^{(0)}}{\upd \rho^2}\right) +\frac{1}{\rho^2}\Sqq^{(1)}W^{(1)}\cos m\theta +2\Srq^{(0)}\left(-\frac{1}{\rho}\frac{\upd W^{(1)}}{\upd \rho} + \frac{1}{\rho^2}W^{(1)}\right) \sin m\theta = 1,
\eeq while the equilibrium of in-plane stresses requires 
\beq
\label{EQ:Balance_O1B}
\frac{1}{\rho}\frac{\partial}{\partial \rho}\left(\rho\Srr^{(0)}\right) +\frac{1}{m}\frac{1}{\rho}\frac{\partial \Srq^{(1)}}{\partial \theta} = 0,\quad\frac{1}{\rho^2}\frac{\partial}{\partial \rho}\left(\rho^2 \Srq^{(0)}\right)+\frac{1}{m}\frac{1}{\rho}\frac{\partial \Sqq^{(1)}}{\partial \theta} = 0.
\eeq As we saw at $O(m)$, a term from bending $\propto\tau^{-2}m^3$ could, in principle, enter the normal force balance \eqref{EQ:Balance_O1A} but is neglected because $\tau=O(m^2)$. Since $\Srq^{(0)}=\Srq^{(0)}(\rho)$, equation \eqref{EQ:Balance_O1A} reveals that the RHS can be balanced only by the first term of the LHS, i.e.
\begin{align}\label{EQ:srr0}
\Srr^{(0)}\left(1-\frac{\upd^2 W^{(0)}}{\upd \rho^2}\right)=1.
\end{align}
Moreover, we find that $\Sqq^{(1)}(\rho,\theta)=0$ since no other terms $\propto \cos m\theta$ are present in the normal force balance equation. When this latter result is used in the second equation of \eqref{EQ:Balance_O1B}, we see immediately that $\Srq^{(0)} (\rho) =A_1/\rho^2=0$, where the constant $A_1$ vanishes on requiring the shear stress to be continuous across the unwrinkled/wrinkled inner interface (noting that the unwrinkled domain is axisymmetric). 
The result, equation \eqref{EQ:srr0}, is consistent with the tension-field approach for the shape of a wrinkled pressurized shell adopted previously \cite{Vella2015_epl}. Moreover, since $\Srr^{(0)}$ is only radius dependent, the first equation of \eqref{EQ:Balance_O1B} then shows that $\Srq^{(1)}=\Srq^{(1)}\left(\rho\right)$ (recall that $\Srq$ must remain $2\pi$-periodic). The first equation of \eqref{EQ:Balance_O1B} then gives  the classic tension-field result $\srr^{(0)}\propto1/\rho$.

Recalling that $\Sqq^{(0)}=0$, the $O(1)$ axisymmetric and asymmetric expansions of the stress-strain relation \eqref{EQ:ett} read, respectively
\beq
\frac{U_r^{(0)}}{\rho} + \frac{\left(W^{(1)}\right)^2}{4\rho^2} + W^{(0)} = -\nu \Srr^{(0)}(\rho)
\eeq
and
\beq\label{EQ:ett_O1_asym}
\frac{1}{\rho}\frac{1}{m}\frac{\partial U_{\theta}^{(1)}}{\partial \theta} - \frac{\left(W^{(1)}\right)^2}{4\rho^2} \cos 2 m \theta = 0.
\eeq
From the first of these, we find the so-called slaving condition \cite{Davidovitch2011}
\beq
\frac{\left(W^{(1)}\right)^2}{4\rho^2} =-\nu\sigma_{rr}^{(0)} - \frac{U^{(0)}_r}{\rho} - W^{(0)}.
\label{eqn:slaving}
\eeq 
Equation \eqref{EQ:ett_O1_asym} then allows us to determine $U_\theta^{(1)}$, though this is of less interest here. Note that, in contrast to the NT analysis, the wrinkle shape, $ W^{(1)}$, is fixed in the FT expansion: a certain amount of length of material has to be `wasted' by wrinkling and the product of the amplitude and the wavenumber is fixed by this condition. Recalling that $\Srq^{(0)}=0$, the $\theta$-dependent expansion of \eqref{EQ:ert} at this order reveals
\begin{align}
\frac{1}{2 \rho}\frac{1}{m}\frac{\partial U_{r}^{(1)}}{\partial \theta} - \frac{1}{2 \rho} \frac{\upd W^{(0)}}{\upd \rho} W^{(1)} \sin m\theta = 0,
\end{align}
so that
\begin{align}\label{EQ:Ur1}
U_r^{(1)} =  - \frac{\upd W^{(0)}}{\upd \rho}  W^{(1)} \cos m\theta,
\end{align}
whereas the $\theta$-independent expansion shows
\begin{align}
\frac{1}{2}\frac{\partial U_{\theta}^{(0)}}{\partial \rho} - \frac{U_{\theta}^{(0)}}{2 \rho} = 0,
\end{align}
and hence that $U_{\theta}^{(0)} =  \rho C_1$, for  $C_1$ a constant. For the hoop displacement $U_{\theta}^{(0)}(\rho)$, to match the axisymmetric state outside the wrinkled region, we must have that $C_1=0$, and hence that $U_{\theta}^{(0)}(\rho)= 0$.

\subsubsection{Order $m^{-1}$}
Assuming $\Sqq^{(1)}=0$ from the results of the previous order, the force balance at $O(1/m)$ reveals that
\begin{align}
\tau^{-2} m^4 \frac{W^{(1)}}{\rho^4}\cos m\theta &+\Srr^{(1)}\left(1-\frac{\upd^2 W^{(0)}}{\upd \rho^2}\right) -\Srr^{(0)}\frac{\upd^2 W^{(1)}}{\upd \rho^2}\cos m\theta +\frac{1}{\rho^2}\Sqq^{(2)}W^{(1)}\cos m\theta \nonumber\\
&+2\Srq^{(1)} \left(-\frac{1}{\rho}\frac{\upd W^{(1)}}{\upd \rho}+\frac{1}{\rho^2}W^{(1)}\right)\sin m\theta = 0.\label{EQ:Balance_O1/mA}
\end{align} Note that the term due to the bending stiffness (terms proportional to $\tau^{-2}$) is finally included at this order since energy minimization led us to expect $\tau^{-2}m^4=O(1)$. The equilibrium of stresses in the other two directions gives 
\beq
\frac{1}{\rho}\frac{\partial}{\partial \rho}\left(\rho\Srr^{(1)}\right) +\frac{1}{m}\frac{1}{\rho}\frac{\partial \Srq^{(2)}}{\partial \theta} = 0,\quad
\frac{1}{\rho^2}\frac{\partial}{\partial \rho}\left(\rho^2\Srq^{(1)}\right) +\frac{1}{m}\frac{1}{\rho}\frac{\partial \Sqq^{(2)}}{\partial \theta} = 0.\label{EQ:Balance_O1/mB}
\eeq

Recalling that $\Srq^{(1)}=\Srq^{(1)}(\rho)$, we see from the second equation of \eqref{EQ:Balance_O1/mB} that $\Sqq^{(2)}(\rho,\theta)=\Sqq^{(2)}(\rho)$ (to ensure $2\pi$-periodicity). We therefore see that the first non-vanishing component of the hoop stress is axisymmetric, as noted previously \cite{Davidovitch2011}. We also find that $\Srq^{(1)}\left(\rho\right)=A_2/\rho^2$, where $A_2$ vanishes upon requiring the stress to vanish at the edges of the wrinkled region.

To say more about wrinkling requires a more detailed analysis of \eqref{EQ:Balance_O1/mA}, which in turn requires that we determine more precisely $\Srr^{(1)}$. We can note that \eqref{EQ:Balance_O1/mA} suggests that $\Srr^{(1)}\propto \cos m\theta$ and so we write
\begin{align}\label{EQ:srr1}
\Srr^{(1)} = g(\rho) \cos m\theta
\end{align} 
where the radial function $g(\rho)$ has yet to be identified. We may then write \eqref{EQ:Balance_O1/mA} as
\begin{align}\label{EQ:VerticalBalance_1overm}
\tau^{-2} \frac{m^4}{\rho^4} W^{(1)} +g(\rho)\left(1-\frac{\upd^2 W^{(0)}}{\upd \rho^2}\right) -\sigma_{rr}^{(0)}\frac{\upd^2 W^{(1)}}{\upd \rho^2}=-\frac{1}{\rho^2}\sigma_{\theta\theta}^{(2)}w^{(1)}.
\end{align}
This has been written in such a way that the right hand side shows the destabilizing hoop stress (the compression that leads to wrinkling), while the left hand side contains the various restoring forces, including the effects of the bending stiffness (first term), a geometrical stiffness due to the curvature of the shell (second term) \cite{Paulsen2016_pnas}, and a tensional stiffness due to the tension along the wrinkles (third term) \cite{Cerda2003_prl}. To make further progress, therefore, we must determine $g(\rho)$; to do so we note that, recalling $\Sqq^{(1)}=0$ and equation \eqref{EQ:Ur1}, the only $\theta$-dependent terms in the $O(1/m)$ expansion of \eqref{EQ:err} are
\begin{align}
\frac{\partial U_r^{(1)}}{\partial \rho} + \frac{\upd W^{(0)}}{\upd \rho}\frac{\upd W^{(1)}}{\upd \rho} \cos m\theta + W^{(1)} \cos m\theta = \Srr^{(1)}= g(\rho)\cos m\theta,
\end{align} 
immediately giving
\begin{align}\label{EQ:G}
g(\rho)=W^{(1)}\left(-\frac{\upd^2 W^{(0)}}{\upd \rho^2}+1\right).
\end{align}
We therefore may rewrite \eqref{EQ:VerticalBalance_1overm} as 
\begin{align}\label{EQ:VerticalBalance_1overmP}
\tau^{-2} \frac{m^4}{\rho^4}  +\left(1-\frac{\upd^2 W^{(0)}}{\upd \rho^2}\right)^2 -\Srr^{(0)}\frac{\frac{\upd^2 W^{(1)}}{\upd \rho^2}}{W^{(1)}}=-\frac{1}{\rho^2}\Sqq^{(2)}.
\end{align} Here we note that the second term on the LHS of \eqref{EQ:VerticalBalance_1overmP} may be written in dimensional terms as
\beq
Y\left(\frac{1}{R}-\frac{\upd^2 w^{(0)}}{\upd r^2}\right)^2;
\eeq this term is a generalization of  the curvature-induced stiffness introduced recently \cite{Hohlfeld2015_pre,Paulsen2016_pnas} for naturally flat sheets, i.e.~$1/R=0$, to objects with an intrinsic curvature.

\subsection{Spatial variation of the wrinkle number}

Equation \eqref{EQ:VerticalBalance_1overmP} is the key result with which we can begin to understand the wrinkle patterns displayed by our ABAQUS simulations. Having generalized the effect of wrinkle curvature studied by Paulsen \emph{et al.} \cite{Paulsen2016_pnas} to include the effect of natural curvature \emph{together} with wrinkle curvature, we shall follow their study to develop a prediction for the wrinkle number.

For any value of the wrinkle number $m$, \eqref{EQ:VerticalBalance_1overmP} gives a solution for the value of $\Sqq^{(2)}(\rho)$ that will give that wrinkle number. However, in the FT approach, we instead ask the question: what value of $m$ minimizes the energy associated with this deformation? In what follows, we assume that the wrinkle number $m(\rho)$ varies only slowly with $\rho$, and so we treat $m(\rho)$ as the value of $m$ that minimizes the energy locally.

Considering the general expression for the elastic energy of the wrinkled state,  \eqref{EQ:energy_functional_general}, we note that the LHS of \eqref{EQ:VerticalBalance_1overmP} corresponds to the functions  
\beq
f_B(\rho)=\frac{1}{\rho^4},\quad
f_K(\rho)=\left(1-\frac{\upd^2 W^{(0)}}{\upd \rho^2}\right)^2 +\Srr^{(0)}\frac{\left(\frac{\upd W^{(1)}}{\upd \rho}\right)^2}{{W^{(1)}}^2}.
\label{eqn:fBfK}
\eeq
Since the slaving condition \eqref{eqn:slaving} fixes $W^{(1)}(\rho)$,  the energy in \eqref{EQ:energy_functional_general} is to be minimized over the wrinkle number $m$, with $f_B(\rho)$ and $f_K(\rho)$ as in \eqref{eqn:fBfK}. Assuming that the wrinkle number varies only slowly with radial position $\rho$ (so that the wrinkle number may be determined locally), we find that
\beq
m(\rho)=\tau^{1/2}\rho(\Kcurv+\Ktens)^{1/4},
\label{eqn:localMlaw}
\eeq where
\beq
\Kcurv=\left(1-\frac{\upd^2 W^{(0)}}{\upd \rho^2}\right)^2
\label{eqn:Kcurv}
\eeq is the curvature--induced stiffness due to the difference between the natural curvature of the shell and the curvature along the wrinkles  caused by indentation, while
\beq
\Ktens=\Srr^{(0)}\frac{\left(\frac{\upd W^{(1)}}{\upd \rho}\right)^2}{{W^{(1)}}^2}
\label{eqn:Ktens}
\eeq is the stiffness due to the tension along the wrinkles \cite{Cerda2003_prl}. Eqn \eqref{eqn:localMlaw} is the appropriate expression of the local $\lambda$-law introduced by Paulsen \emph{et al.}~\cite{Paulsen2016_pnas} in a setting with circular symmetry, where it is more natural to talk about wavenumber than wavelength. There are two important differences here with the local $\lambda$-law as previously expressed: first, our expression for $\Kcurv$ \eqref{eqn:Kcurv} includes the effect of natural curvature (rather than imposed curvature, as studied previously \cite{Hohlfeld2015_pre,Paulsen2016_pnas}). Second, our problem does not contain a stiffness due to an elastic foundation, though this could be added to \eqref{eqn:localMlaw} in the obvious way.

To make the final step, we require the base state of the wrinkled shell $W^{(0)}$; the calculation of $W^{(0)}$ using tension field theory, is outlined by Vella \emph{et al.} \cite{Vella2015_epl}. In general, a multipoint boundary value problem must be solved numerically to find $W^{(0)}(\rho)$. However, in the limit of $\tdelta\gg1$ Vella \emph{et al.} \cite{Vella2015_epl} showed that wrinkles occupy an increasing portion of the shell, $1.17\tdelta^{-1/2}\leq\rho\leq 1.19013\tdelta^{1/2}$ and that, within this region, the base state of the shell is given by
\beq
W^{(0)}(\rho)\sim\tdelta\left[-1+0.963\frac{\rho}{\tdelta^{1/2}}+\tfrac{1}{2}\left(\frac{\rho}{\tdelta^{1/2}}\right)^2-0.507\left(\frac{\rho}{\tdelta^{1/2}}\right)^3\right].
\label{eqn:TFTprof}
\eeq 

Using this basic shape, together with \eqref{eqn:Kcurv}, it is then easy to see that
\beq
\Kcurv\approx 9.242\rho^2\tdelta^{-1}.
\eeq 

The slaving condition \eqref{eqn:slaving} can be used with the tension-field profile for $W^{(0)}(\rho)$, \eqref{eqn:TFTprof}, to determine the wrinkle amplitude $W^{(1)}(\rho)$; this has been calculated previously \cite{Vella2015_epl}. Using this previous result, we find that in the limit $\tdelta\gg1$
\beq
\Ktens\approx 0.329 \frac{\tdelta^{1/2}}{\rho^3} \left(\frac{1.416 \tdelta-3 \rho^2}{1.416 \tdelta-\rho^2}\right)^2.
\eeq It is clear that this expression diverges both at the indenter position, $\rho\to0$, and at the edge of the wrinkled zone, $\rho\approx1.19\tdelta^{1/2}$. Our assumption in the analysis leading to \eqref{eqn:localMlaw} was that variations in $m$ occur only slowly with $\rho$; while this remains plausible as $\rho\to0$ (since then $m\sim \rho^{1/4}$), it is clearly not the case as $\rho\to1.19\tdelta^{1/2}$. In fact, near the edge of the wrinkles there is a spatial boundary layer in which this divergence is resolved; we discuss this further in \S\ref{sec:conclude} but for now proceed by approximating $\Ktens$ as
\beq
\Ktens\approx 0.329 \tdelta^{1/2}\rho^{-3}.
\eeq
Combining the expression for the two stiffnesses with \eqref{eqn:localMlaw} the radial variation of the wavenumber is then approximately described, for $\tdelta\gg1$, by
\beq
m(\rho)=\tau^{1/2}\rho\left(9.242\frac{\rho^2}{\tdelta}+\frac{0.329\tdelta^{1/2}}{\rho^3}\right)^{1/4}.
\label{eqn:radialM}
\eeq
We now compare this result with the results of our Finite Element analysis.

\section{Comparison between `Far from threshold' analysis and ABAQUS\label{sec:compare}}

The preceding analysis resulted in a (relatively) simple prediction for the spatial variation of the number of wrinkles, $m(\rho)$, in the limit of very large indentation depths $\tdelta\gg1$. The scaling behaviour $m\sim \tau^{1/2}$ has been predicted previously \cite{Vella2011_prl,Vella2015_epl}. However, we emphasize that the dependence on indentation depth $\tdelta$ and radial position, $\rho$, are novel. Furthermore, the presence of two different stiffnesses, $\Kcurv$ and $\Ktens$, together with the very different behaviours of these two stiffnesses with radial position and indentation depth, $\tdelta$, mean that different scaling laws may hold in different regions of the shell. In particular, we see that the two tensions balance one another when
\beq
\rho\sim\bar{\rho}=0.513\tdelta^{3/10}.
\eeq Roughly speaking, then, we expect that for $\rho\lesssim\bar{\rho}$, $\Ktens$ dominates and the wrinkles are `tensional'; for $\rho\gtrsim\bar{\rho}$, it is $\Kcurv$ that dominates and the wrinkles are `curvature-dominated'. Finally, we note that wrinkles only occupy the region $\rho\lesssim1.19\tdelta^{1/2}$ and that the wrinkled shell adopts a universal shape, given by \eqref{eqn:TFTprof}, in this case. It is therefore natural to rescale $\rho$ to exploit this universal shape, which we do by letting  $\xi=\rho/\tdelta^{1/2}=r/(\delta R)^{1/2}$. In this rescaled coordinate, wrinkling in the majority of the shell is dominated by $\Kcurv$ (since $\bar{\rho}/\tdelta^{1/2}\to 0$ as $\tdelta\to\infty$) and so we have
\beq
m(\xi)\left(\frac{t}{\delta}\right)^{1/2}\approx1.743\left[12(1-\nu^2)\right]^{1/4}\xi^{3/2},
\label{EQ:FFT_normalization}
\eeq  where the dependence on Poisson ratio $\nu$ emerges from the ratio of the bending and stretching moduli, $B/Et=t^2/12(1-\nu^2)$. We note that written in this way, the wrinkle number is independent of the modulus and the pressure, depending only on the geometrical properties of the shell and the deformation. This result is somewhat surprising but has also been observed in a thin membrane floating on a liquid \cite{Paulsen2016_pnas}.

A particular instance of \eqref{EQ:FFT_normalization} is that the wrinkle number at the edge, $m=\medge$, may be found by substituting $\xi=1.19$, to give
\beq
\medge\approx 2.26\tau^{1/2}\tdelta^{1/2}.
\eeq We note that this corresponds to a constant wavelength at the wrinkles' edge as indentation progresses. Results for other radial positions are summarized in table \ref{TAB:m_FFT_scaling}. We note in particular that when viewed at a fixed radial position, $\rho_{\mathrm{fix}}$, the number of wrinkles observed initially decreases (according to \eqref{EQ:FFT_normalization}), while for very large indentation depths it will increase once $\rho_{\mathrm{fix}}<\bar{\rho}$.

\begin{table}[!h]
\centering
\caption{Expected scaling laws for the wavenumber $m$ observed in the different (spatial) regions of an indented, highly pressurized elastic shell (i.e.~$\tau\gg1$, $\tdelta\gg1$).}
\label{TAB:m_FFT_scaling}
\centering
\begin{tabular}{lll}
\hline
Region &Dominant Stiffness &Wavenumber  \\
\hline
$\rho \lesssim \bar{\rho}$ & $\Ktens$ &$m  \approx 0.75  \tau^{1/2} \tdelta^{1/8} \rho^{1/4}$  \\
$\rho \approx \bar{\rho}$ &$\Ktens\approx \Kcurv$ & $m  \approx 0.76 \tau^{1/2} \tdelta^{1/5} $ \\
$\rho\gtrsim \bar{\rho}$ &$\Kcurv$ &$m  \approx 1.73 \tau^{1/2}\tdelta^{-1/4}\rho^{3/2}  $  \\
$\rho = \rho_{O}$ &$\Kcurv$ & $m  \approx 2.26\tau^{1/2}\tdelta^{1/2}$  \\\hline
\end{tabular}
\vspace*{-4pt}
\end{table}

\begin{figure}[h]
\centering
\centering\includegraphics[width=13.7cm]{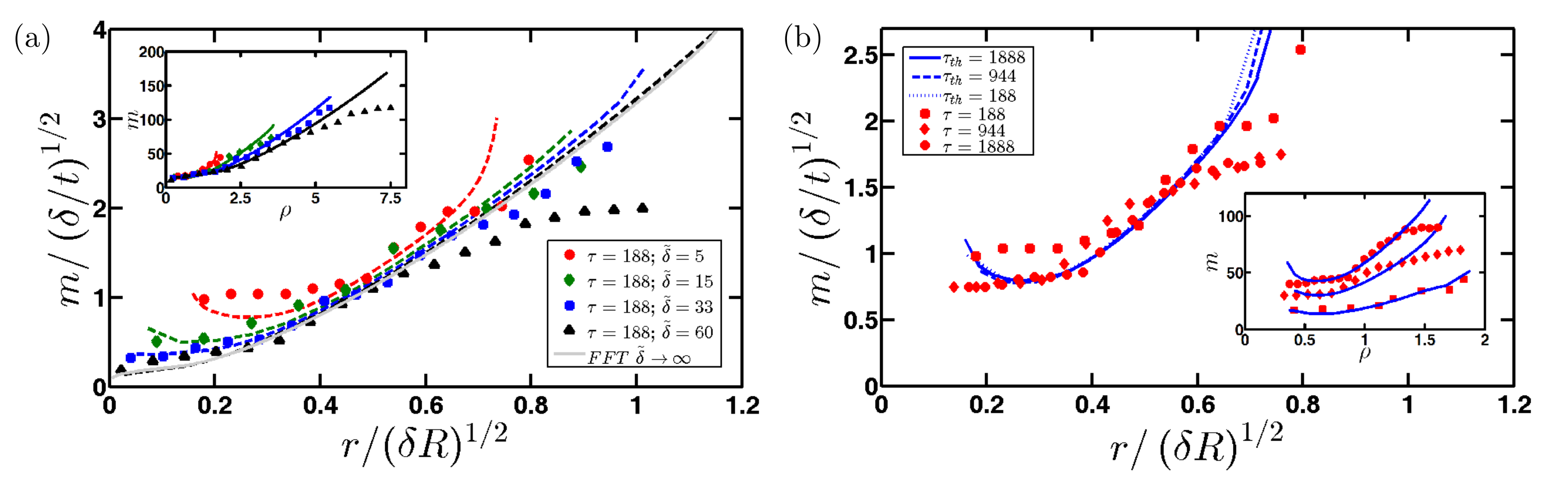}
\caption{Radial variation in the wrinkle number. (a) Effect of indentation depth $\tdelta$ on the local wavenumber $m$ for fixed $\tau=188$. The solid grey curve shows the analytical expression  \eqref{EQ:FFT_normalization}, valid for $\tdelta\gg1$, as discussed in the text, while dashed curves indicate the prediction of the FT analysis when the functions $W^{(0)}(\rho)$ and $W^{(1)}(\rho)$ are computed numerically for given finite $\tdelta$. (b) Effect of pressurization $\tau$ on the local wavenumber $m$ observed at a fixed indentation depth $\tdelta=5$. Again, blue curves indicate the FT prediction based on the functions $W^{(0)}(\rho)$ and $W^{(1)}(\rho)$  computed numerically for  finite $\tdelta$. In both (a) and (b), the raw value of $m$ is plotted against the  dimensionless radial coordinate $\rho$ in the insets.} \label{FIG:normalization}
\end{figure}

Figure \ref{FIG:normalization} presents a comparison for the wrinkle numbers predicted  theoretically and those observed in our finite element simulations. In particular, figure ~\ref{FIG:normalization}(a) investigates the effect of varying $\tdelta$ whilst holding $\tau=188$ fixed, while figure ~\ref{FIG:normalization}(b) investigates the effect of varying $\tau$ whilst holding $\tdelta=5$ fixed. In each case, the raw dimensionless data showing the variation of $m$ against the radial coordinate $\rho$ is presented in the inset, while the main figure shows the  data normalized according to \eqref{EQ:FFT_normalization}.

The agreement between theory and numerics shown in figure~\ref{FIG:normalization} is generally very good. However, this comparison also highlights that the agreement begins to break down at the largest indentation depths $\tdelta\gtrsim33$,  possibly in a manner similar to the crumpling observed in a film on a liquid drop \cite{King2012_pnas}. Indeed, we note that at the external boundary of the wrinkled zone, the simulation results shown in figure~\ref{FIG:Deformeds}(d) are highly reminiscent of the crumples observed in the sheet-on-drop problem (see for example figure~3C of King \emph{et al.}~\cite{King2012_pnas}).

\begin{figure}[h!]
\centering\includegraphics[width=13.7cm]{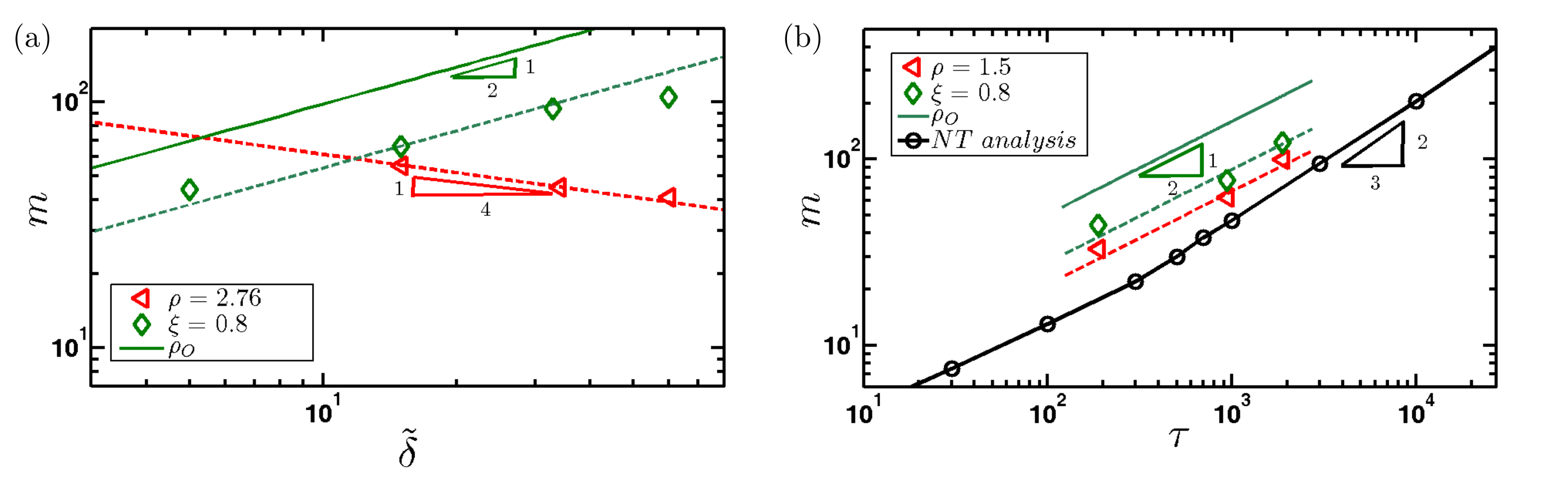}
\caption{Wrinkle wavenumber $m$ against (a) the dimensionless indentation depth $\tdelta$ (at fixed $\tau = 188$) and (b) the dimensionless pressurization $\tau$ (at fixed $\tdelta = 5$), at different spatial positions within the shell. Three relevant results for the domain where $K_{curv}$ dominates are shown: outer boundary $\rho_O$ (solid green line); fixed dimensionless radial position $\rho=r/\left(\delta R\right)^{1/2}$ (dashed red line); fixed universal position $\xi=\rho/\tdelta^{1/2}$ (dashed green line). Points show the data extracted from the finite element simulations while the relevant lines show the predicted scaling laws (including prefactors).}\label{FIG:all_scaling} 
\end{figure}

Table \ref{TAB:m_FFT_scaling}  summarizes four different scaling behaviours, depending on which region of the shell we examine. Figure \ref{FIG:all_scaling} shows tests of predictions for the region $\rho>\bar{\rho}$ in which $\Kcurv$ dominates (as $\tdelta$ increases, this region fills an increasing proportion of the shell). We emphasize that apparently complex behaviour can observed as  $\tdelta$ increases: for a given radial coordinate $\rho$, the wrinkle wavenumber decreases with $\tdelta$ but if we instead focus on a fixed universal position, $\xi=\rho/\tdelta^{1/2}$ fixed, we see that $m$ increases. 
 
Despite the different spatial variations of wrinkle number, we always find that $m \sim \tau^{1/2}$, which is the typical scaling found in Far from Threshold analyses \cite{Davidovitch2011,Paulsen2016_pnas}. It is particularly important to emphasize that this scaling law is different from the prediction of the Near Threshold analysis,  which yielded $m\sim \tau^{2/3}$ and is reproduced as the black curve in Figure \ref{FIG:all_scaling}(b). Our finite element simulations show both the NT regime and the FT regime; we believe that this is the first time both of these have been recovered in finite element simulations, though both regions have been studied in the theoretical analysis of Davidovitch {\it et al.} \cite{Davidovitch2012_pre}. Indeed, from a numerical point of view, the comparison provided in Taylor {\it et al.} \cite{Taylor2015_jmps}  suggests that different numerical methods are better suited to either the NT or the FT regimes, not both. We also note that, in our results, it seems that the transition between the two regimes is not clear-cut.

It is interesting to note that the scaling of the wavenumber with $\tau$, i.e.~$m\sim\tau^{1/2}$, has already been suggested in two previous works \cite{Vella2011_prl,Vella2015_epl}.  Vella {\it et al.} \cite{Vella2011_prl} used a scaling argument along the lines of \cite{Cerda2003_prl} to explain the wrinkle number observed at onset (the Near threshold behaviour). Such an argument should only be used to determine the FT behaviour, however; the apparently good agreement between the two is, we believe, an accident of the fact that the wrinkle number at onset only reaches the expected $\tau^{2/3}$ scaling when $\tau\ggg1$ while, for the values investigated previously \cite{Vella2011_prl}, the NT behaviour was still transitioning between the $\tau\ll1$ and $\tau\gg1$ regimes. A later consideration of the global energy consistent with the FT approach was given by Vella {\it et al.} \cite{Vella2015_epl}. This result predicted some variation in wrinkle wavenumber with indentation depth (though did not observe such variation), but did not take into account the spatial variation of the wrinkle pattern. As we have seen in this work, the radial variation means that  there are no simple scalings for the evolution of the wrinkle pattern with applied indentation: one can see that the wrinkle number increases or decreases with indentation depth depending on whether one remains at fixed  radial position, or moves with the edge of the wrinkled zone.

\section{Conclusion\label{sec:conclude}}

In this paper, we have presented a combined numerical and asymptotic study of the well-developed wrinkling of a pressurized shell subject to point indentation. We have shown that the scaling for the wrinkle number depends on the dimensionless pressurization $\tau$ according to $m\sim \tau^{1/2}$; a key result of our analysis is that this `Far from threshold' scaling is different to the scaling $m\sim \tau^{2/3}$ that is observed at the onset of instability. Furthermore, we have shown that at a fixed indentation depth, the wrinkle number varies spatially within the shell, according to which of two effective stiffnesses (one dominated by tension, the  other by curvature) is more significant. Finally, we generalized the concept of a curvature-induced stiffness  proposed by Paulsen et al. \cite{Paulsen2016_pnas} to situations in which this curvature occurs together with an intrinsic curvature.

Generally speaking, we found very good agreement between finite element simulations and our asymptotic analysis. Nevertheless, a number of features of our results warrant further investigation. One such feature is the discrepancy found for very large indentations at the outer boundary of the wrinkled area (the crumples shown in the snapshots of deformed shapes shown in figure~\ref{FIG:Deformeds}(d), together with the decrease in wrinkle number observed there, compared with the FT prediction). We note that while strikingly similar crumples have been observed experimentally in related systems (see for example \cite{King2012_pnas}), one possible confounding factor here is our use of shallow shell theory. Shallow shell theory requires that the horizontal scale of the deformation is much smaller than the radius of curvature of the shell itself. In the largest indentation depths probed here ($\tdelta\approx60$, $\lp\approx5\mathrm{~cm}$) the extent of the wrinkled region, $r_O\approx 0.5\mathrm{~m}$, which is close to the scale of the shell itself (recall that here $L=0.8\mathrm{~m}$). As a result, future work should focus on understanding whether this drop in wrinkle number at large indentation depths is a finite size effect, or a robust feature due to crumpling.
The numerical simulations also show a second interesting behaviour. At large indentations, for example $\tdelta=33$ or $\tdelta=60$ in Figure \ref{FIG:MinPrincipal}, some wrinkles are visually more apparent than others. This is reminiscent of the wrinkle-to-fold transition that has already been noted in related problems, such as the confinement of a thin film over a liquid substrate \cite{Brau2013_sm}. This transition is still to be fully understood in the relatively simple two-dimensional setting of a sheet on a liquid bath, and so is beyond the scope of the present work; nevertheless this remains an interesting possibility for future work.

\begin{figure}[!h]
\centering\includegraphics[width=5.0in]{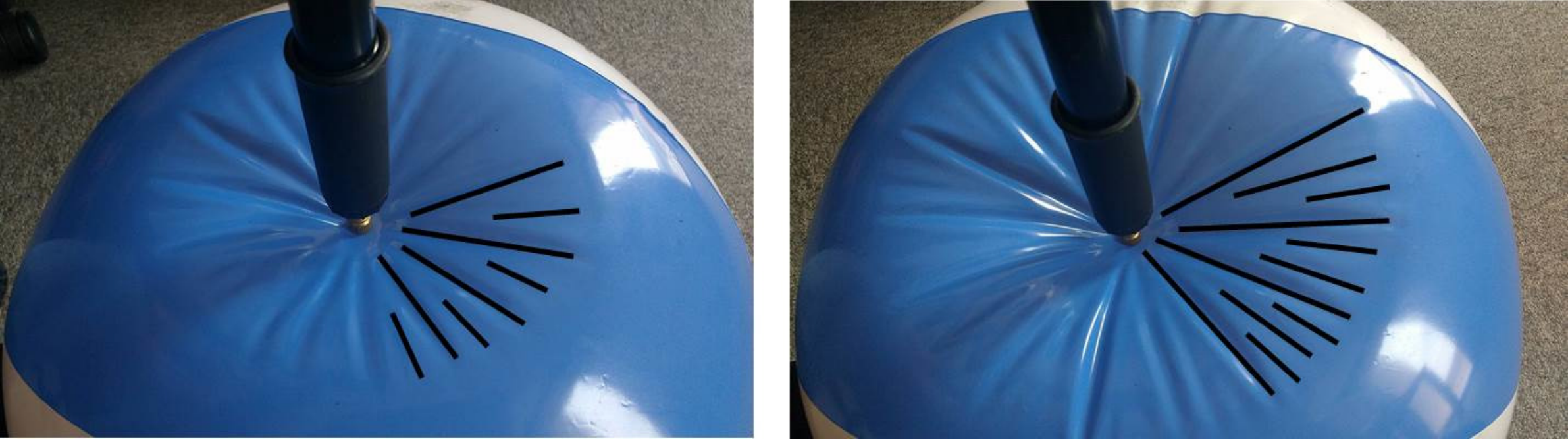}
\caption{Indentation of a pressurized beachball. Using two different indentation depths, we find qualitative support for the results of the analysis presented in this paper: firstly, the wrinkling pattern is affected by the indentation depth (compare left versus right) while the number of wrinkles increases with horizontal distance from the indenter. (To highlight the troughs of the wrinkles, especially the appearance of new wrinkles, black line segments have been superimposed on a portion of the images.) Secondly, we see no signs of the wavenumber divergence that would be expected at the edge of the wrinkled portion, $\rho\to1.19\tdelta^{1/2}$, if $\Ktens\to\infty$ there. Finally, the wavelength at this outer edge appears to be approximately constant.}
\label{FIG:balloon}
\end{figure}

Another pressing theoretical concern exposed by our analysis is the apparent divergence of the tensional stiffness at the edge of the wrinkled zone, $\rho\approx1.19\tdelta^{1/2}$. 
Simple experiments with a beachball (see figure~\ref{FIG:balloon}) confirm our finding from finite element simulations that the wrinkle number does not diverge at the edge of the wrinkled zone and so we rule out the presence of a cascade of wrinkles as seen at the boundaries of wrinkled zones in related problems \cite{Huang2010,Bella2016}. 
From \eqref{eqn:Ktens} it is clear that the divergence in wrinkle number is caused by the divergence in $\Ktens$ that occurs when $W^{(1)}\to0$ with $\upd W^{(1)}/\upd \rho $ finite. In reality, this divergence is regularized by accounting for the term $\bigl(\upd W^{(1)}/\upd \rho\bigr)^2$ in the computation of the slaving condition, as shown in a similar problem \cite{Davidovitch2012_pre}. (This term is neglected in our analysis because of our assumption that radial derivatives are $O(1)$ while azimuthal derivatives are $O(m)$. This assumption breaks down in a boundary layer near the outer edge of the wrinkles.) For our purposes, we note merely that the existence of this smoothing out justifies our assumption that $\Ktens$ can be neglected near the edge of the wrinkled zone (as was also observed in the Lam\'{e} problem \cite{Taylor2015_jmps}). In turn, our neglect of $\Ktens$ at the edge leads to the prediction of a constant wrinkle wavelength at the edge of the wrinkled zone, in qualitative agreement with figure \ref{FIG:balloon}.  

Finally, we note that much of the experimental interest in understanding well-developed wrinkle patterns, particularly the wrinkle number, has come from the desire to use wrinkles as a metrological assay for determining the properties of thin sheets \cite{Huang2007,Knoche2013}. Our results here, particularly the prediction of a curvature-induced stiffness that depends on both natural and imposed curvature, may lead to a reassessment of previous experimental work that used the  wrinkling of deflated bubbles to infer the 2D modulus of complex interfaces \cite{Knoche2013}. Our observartion of a spatially variable wrinkle wavelength also suggest a possible mechanism for `chirp' in photonics applications.

\vskip6pt


\section*{Acknowledgment}

The research leading to these results has received funding from the European Research Council under the European Union's Horizon 2020 Programme / ERC Grant Agreement no.~637334 (DV). We are grateful to Benny Davidovitch for many discussions about the wrinkling of pressurized shells, and for thoughtful comments on an earlier draft of this paper.  Data associated with this paper may be found at \texttt{https://doi.org/10.5287/bodleian:DOeqDMwqG}.

\end{document}